\documentstyle[preprint,aps]{revtex}
\begin{document}
\draft
\title{Theory of a Scanning Tunneling Microscope with a
Two-Protrusion Tip}
\author{Michael E. Flatt\'{e}$^1$, Jeff M. Byers$^2$}
\address{$^1$ Department of Physics and Astronomy, University of
Iowa, Iowa City, Iowa 52242}
\address{$^2$ Naval Research Laboratory, Washington D.C. 20375 }
\date{July 21, 1995}
\maketitle
\begin{abstract}
We consider a scanning tunneling microscope
(STM) such that tunneling occurs through two atomically sharp
protrusions on its tip.
When the two protrusions are separated by at least
several atomic spacings, the differential conductance of this STM
depends on the electronic transport in the
sample between the protrusions.
Such two-protrusion tips commonly occur
during STM tip preparation. We explore possible applications to probing
dynamical impurity potentials on a metallic surface and local
transport in an anisotropic superconductor.
\end{abstract}
\vfill\eject
\narrowtext

Scanning tunneling microscopy (STM)
enables the characterization
of materials on the atomic scale through
measurements of the local density of states (LDOS).
Recently, in a series of STM experiments on the Cu (111) surface,
the local transport properties of
electrons in a Shockley surface state were probed through
their influence on the LDOS around an Fe impurity\cite{1}.
A similar experiment has been proposed for measuring the
transport properties of a high-temperature
superconductor\cite{2}.
These experiments detect the reflection from the impurity of electrons injected
by the STM. The
spatial resolution of these experiments is sub-Angstrom.
Properties which might be determined from these types of
measurements, but could not be probed by an STM
measurement on the homogeneous sample,
include the {\it angularly-resolved}
dispersion relations and mean free path, as well as the  density of
states as a function of energy {\it and} momentum.
In the Cu experiments the dispersion relation was
measured, no evidence for density-of-states anisotropy
with angle was found, and the mean free path was too large
to detect.

A stronger signal would result from an independent method of injecting
electrons at a site (by another contact) and detecting them elsewhere
with an STM.  The possibilities for this have been explored in
recent work\cite{3,4} and will be referred to as the
two-contact experiment.
A related technique, applicable to transport on a longer length scale
($\mu$m), uses a laser to create nonequilibrium quasiparticles and a
point contact to detect them\cite{5}.

The STM is also sensitive to the spectroscopic properties of
impurities\cite{6,7}. However an STM averages over the
fluctuating part of an impurity potential, such as that of a free
orbital moment, making it difficult to observe.
Such free moments are of great interest in part because their
interaction with conduction electrons may produce a Kondo resonance.

We suggest in this Letter an experiment which should
provide detailed angular information about fluctuating impurity
potentials {\it and} probe transport on a homogeneous sample.
The apparatus, shown schematically in Figure 1, would consist of a
spatially-extended STM tip with two protrusions each ending in a single
atom. Although the
case of STM tips with tunneling at more than one place has been
considered before\cite{8}, that work was concerned with
tips ending in {\it clusters} with substantial tunneling through more
than one atom. The images obtained from these STM tips tended to
be blurry and less useful than those from single-atom-terminated
contacts. Here we propose a tip with two atomically-sharp protrusions,
and will demonstrate that new information is obtainable when these
protrusions are separated by more than 10\AA. In contrast to the
difficulties associated with arranging two independent {\it contacts}
in close proximity, two-{\it protrusion} tips are often created by
chance during tip preparation, with tip separations up to
$1000$\AA\cite{9}.
The interference between the two protrusions influences
tunneling conductances at a lower order in the tunneling matrix
elements than the two-contact experiment. The two-protrusion
experiment, therefore, should be easier to construct and have greater
signal than the two-contact experiment.

For separations of $10$\AA$-100$\AA\ a two-protrusion tip
would be useful for probing the angular structure of a free
moment. The differential conductivity depends
on the angle-resolved amplitude for
electrons to scatter from the impurity. A measurement with a
single-protrusion STM tip merely measures backscattering.
For an impurity state {\it
fixed} relative to the lattice orientation by the crystal
field, backscattering is sufficient to determine the
impurity's angular structure; therefore the single-protrusion
measurement provides as much
information as the two-protrusion one. However, for
identifying the angular structure of a {\it free} moment,
the two-protrusion
tip is superior to the single-protrusion tip. A two-contact
experiment could in principle measure this angular structure
as well\cite{4}, but positioning two tips within $100$\AA\
of each other would be extremely difficult.

On a homogeneous sample the transport quantities of interest
would determine the desired
separation of protrusions on the STM tip.
Measurements of quantities with long length scales
($100$\AA$-1000$\AA) such as mean free paths, transitions from ballistic to
diffusive propagation, low-$T_c$ superconductors' coherence lengths,
charge-density-wave correlation lengths, and angularly anisotropic
density-of-states effects\cite{2} would most benefit
from the increased signal of the two-protrusion configuration
relative to the two-contact configuration. It is also at
these distances that the overlapping interference of
other impurities on a surface would complicate a
measurement performed with a single-protrusion STM around an
impurity. However, for electronic quantities with short
length scales, such as
Fermi wavelengths, the single-protrusion STM
would likely perform the best of the three.

The tunneling Hamiltonian
\begin{equation}
H_T = \int d{\bf x}d{\bf r}\left[
T({\bf x},{\bf r})\psi^\dagger({\bf x}
)\phi({\bf r})+{\rm H.c.}\right],
\end{equation}
where $\psi({\bf x})$ is the field annihilation operator for
an electron at position ${\bf x}$ in the STM tip, and $\phi({\bf
r})$ is the field
annihilation operator for an electron at position ${\bf r}$
in the sample. The electron spin is treated implicitly to
simplify the notation. The
differential conductance of the STM at $T=0K$,  \begin{equation}
{dI\over dV} =
={2e^2\over h}
\int d{\bf x}d{\bf r}d{\bf x'}d{\bf r'}
T({\bf x},{\bf r})T^*({\bf x'},{\bf r'})
{\rm Im}g({\bf x},{\bf x'};0){\rm Im}G({\bf r'},{\bf
r};eV), \end{equation}
where $G$ is the Green
function in the sample and $g$ is the Green function in the STM
tip.

Typically the transfer function $T({\bf x},{\bf r})$ is taken to be localized
near a point in the sample and the tip:
$T({\bf x},{\bf r}) = W\upsilon({\bf x}-{\bf x_o})\upsilon'({\bf
r}-{\bf r_o})$, where the integrals of
$|\upsilon|^2$ and $|\upsilon'|^2$ are unity. With $\upsilon'$ a
highly-localized function the differential conductance is proportional to
\begin{equation}
{\rm Im} \tilde G({\bf r_o},{\bf r_o};eV) = {\rm Im}\int d{\bf r}d{\bf
r'} \upsilon'({\bf
r}-{\bf r_o})\upsilon'^*({\bf r'}-{\bf r_o})G({\bf r},{\bf
r'};eV)\sim {\rm
Im}G({\bf r_o},{\bf r_o};eV).
\end{equation}
When $\upsilon$ is also very localized
the proportionality constant is $e^2|W|^2N(0)/h$, where $N(0)$ is the
density of states in the tip at the Fermi energy. The expression
$dI/dV = e^2|W|^2N(0){\rm Im} G({\bf r_o},{\bf r_o};eV)/h$ is a common
starting point for STM theory\cite{10}.

We model the effect of a single impurity resonance by the
Hamiltonian term
\begin{equation}
H_{I} = \int d{\bf r} \left [  A({\bf r})\phi({\bf r})^\dagger \chi +
A^*({\bf r})\phi({\bf r})\chi^\dagger\right]
\end{equation}
where $\chi$ is the annihilation operator for an electron in
the localized state. For a simple model\cite{11}, $A({\bf
r})\propto \Psi({\bf r})$, the (normalized) wavefunction of the impurity
state. The Green function is then
\begin{eqnarray}
G({\bf r_1},{\bf r_2};eV) =&& \int d{\bf r}d{\bf r'}G({\bf
r_1},{\bf r};eV){A({\bf r})A^*({\bf r'})\over
eV-E_I+i\Gamma}G({\bf r'},{\bf r_2};eV)\label{E1}\\
=&& \int d{\bf r}d{\bf r'}G({\bf
r_1},{\bf r};eV){\Psi({\bf r})\Psi^*({\bf r'})\over \pi
N_s}
G({\bf r'},{\bf r_2};eV),\qquad eV=E_I\label{E2}
\end{eqnarray}
where
$E_I$ and $\Gamma$ are the energy and linewidth of the
impurity state and $N_s$ is the density of states of the sample at
$E_I$. We assume there is no other influence on $\Gamma$ besides
hopping to the extended states. For a $d$-state,
$\Psi({\bf r}) = (2/\xi\sqrt{\pi}){\rm e}^{-r/\xi}
\cos(2\theta_{\bf r}),$
where $\xi$ is the range of the state. For a fixed
$d$-state, $\theta_{\bf r}$ is measured relative to a
crystallographic axis.
Figure 2 shows the
single-protrusion differential conductance in the vicinity
of a fixed impurity $d$-state with $\xi = 2k_V^{-1}$ ($k_V$
is the wavenumber of the electronic state with energy $eV$).  It
shows clear four-fold symmetry.
If the crystal field splitting of the impurity levels is larger than
the temperature and $\Gamma$, the STM can
separately probe each non-degenerate level by adjusting the voltage.

The two-protrusion
experiment becomes more useful than the single-protrusion experiment
when the impurity of interest has a free moment.
Then the $dI/dV$ must be averaged over orientations of
$\theta_{\bf r}$ in Eq.~(\ref{E2}) (and Fig. 2). The two-protrusion
transfer function, however, contains contributions from  both tunneling
sites below the protrusions: $T({\bf x},{\bf r}) =
\sum_i W_i\upsilon_i({\bf x}-{\bf x_i})\upsilon_i'({\bf r}-{\bf
r_i})$, where for simplicity we consider $\upsilon$, $\upsilon'$,
and $W$ to be independent of $i$. The STM current for this system is
\begin{equation} {dI\over dV} =
{2e^2|W|^2\over h}\sum_{\{i,j\}=1}^2{\rm Im}\tilde g({\bf x_i},{\bf
x_j};0){\rm Im}\tilde G({\bf r_j},{\bf r_i};eV) .\label{E3}
 \end{equation}
The $i=j$ terms describe direct tunneling through the two
protrusions, but the $i\ne j$ terms are interference terms between the
two protrusions.
We approximate ${\rm Im}\tilde g({\bf x},{\bf x'};0)$ by
$N(0)\exp(-|{\bf x}-{\bf x'}|/\ell_I)$ where $\ell_I$ is the
inelastic mean free path\cite{12}.
A favorable tip material would have a large $N(0)$ and a
long $\ell_I$.
Figure 3
shows two-protrusion measurements of a $d$-orbital which is free to
move in the plane of the surface.
The $dI/dV$ is plotted as a function of distance from
the impurity $r$ (same for both protrusions) and angle $\theta$
between the two tips.
A single-protrusion
measurement corresponds to $\theta=0$. The four-fold structure of the
$d$-orbital is clearly visible in Figure 3, which should be
thought of as a compilation of results which must come from many different
tips, since for each tip the distance between protrusions is
fixed.

A measurement with a particular tip, with separation
$7.5k_V^{-1}$, is shown in Figure 4.  The
differential conductance through the two protrusions at ${\bf
r_1}=(x,3.75)$ and ${\bf r_2} = (x,-3.75)$ is
compared to the differential conductance of a single protrusion
located at ${\bf r_1}$ with tunneling matrix element $2W$. The
geometry is shown in Figure 1. The
differential conductances for $x=0$ ($\theta=\pi$)
are identical and at large $|x|$
(where $\theta$ is small) the two are very similar. The most
prominent difference is the absence in the two-protrusion $dI/dV$ of
a peak near $x=\pm 3.75$. Since $\theta=\pi/2$ for $x=3.75$, the
direct $i=j$ terms are almost fully cancelled by the $i\ne j$
terms in Eq.~(\ref{E3}). We emphasize there is no need to rotate
the tip assembly to perform this measurement. If the impurity moment
is free, for any tip orientation the geometry of Figure 1 can be
arranged solely by translation of the tip.
The orientation and protrusion separation of the tip can be
identified by analyzing the double image from an impurity or a step
edge.

We now apply Eq.~(\ref{E3}) to transport between the two
protrusions through a $d_{x^2-y^2}$-gapped superconductor.
A $d_{x^2-y^2}$ gap has been proposed\cite{13} for high-$\rm T_c$
superconductors, including $\rm Bi_2Sr_2CaCu_2O_8$.
A very clean surface can be
prepared on $\rm Bi_2Sr_2CaCu_2O_8$ so that
transport is ballistic, and
the mean-free path is
$10^3-10^4$\AA\cite{14} at $4.2$K.
${\rm Bi_2Sr_2CaCu_2O_8}$
is a layered structure with weakly-coupled planes, so we calculate
the Green's functions for a $d_{x^2-y^2}$-gapped
superconductor with a cylindrical Fermi surface\cite{2}.
Figure\ \ref{fig2} shows the position-dependent differential conductance
as a function of $x$ and $y$ for the
$d_{x^2-y^2}$ gap  $\Delta_{{\bf k}} = \Delta_{max} \cos(2\phi_{\bf k})$.
$\phi_{\bf k}$ is the angle that  the momentum ${\bf k}$ makes with the
crystallographic {\it a}-axis. The voltage bias is set well below the
gap maximum ($eV=0.1\Delta_{max}$)  so that the
quasiparticles are only able to propagate in the directions where
$\Delta_{{\bf k}}$ has nodes.

In a heuristic sense, gap anisotropy produces an
angularly-dependent density of states, which can be
qualitatively different at different energies. For a
$d_{x^2-y^2}$ gap, at voltages
much less than the gap quasiparticles can only travel in the
real-space directions roughly parallel to node momenta,
yielding ``channels'' of conductance\cite{3}. At voltages slightly
higher than the gap maximum there are more states for
momenta near the gap maximum, so the channels would appear
rotated by $45^o$. Measurements of gap anisotropy, particularly from
angle-resolved photoemission\cite{15}, are of great current interest
for distinguishing among various theories of high-temperature
superconductivity. Tunneling experiments have an energy resolution
better than an meV, far superior to angle-resolved photoemission.
Again it is not necessary to rotate the
tip assembly since regions of the sample with different orientations
(separated by grain boundaries) could be measured instead.

The signal in this two-protrusion transport
experiment is greater than the two-contact or impurity
configurations because the interference terms in
Eq.~(\ref{E3}) are first-order (proportional to
$|W|^2$). The two-contact experiment relies on a second-order process,
proportional to $|W|^4$\cite{3}. The impurity transport
experiment\cite{2}
relies on a process which is first-order in tunneling, $|W|^2$, and
impurity-scattering $|U|^2$ (the $U$ is the
potential strength),
and thus overall is second order ($|W|^2|U|^2$).

The primary goal of this Letter has been to offer an example of how
a two-protrusion STM can explore the characteristics
of fluctuating impurity potentials and the local transport properties
of a homogeneous sample.
A two-protrusion $dI/dV$ indicates
the angular symmetry of an impurity state in a similar way to how a
differential cross-section of a particle-atom collision indicates the
angular  symmetry of
an atomic state.
Measuring gap anisotropy is
merely one possible application of the ability to directly and in
detail probe small-scale electronic transport in a
homogeneous sample.

We acknowledge useful conversations with M.F. Crommie, C.M. Lieber,
and A. Yazdani. J.M.B. is supported by an NRC Fellowship.

\begin{figure}
\caption{Probe-sample geometry. Two protrusions on a single STM contact
are connected to the same reservoir.  }
\label{fig1}
\end{figure}
\begin{figure}
\caption{Single-protrusion STM differential conductance
for an impurity $d$-state with $\xi=2k_V^{-1}$ in units of
$e^2|W|^2 N(0)N_s/h$, where $N(0)$ and $N_s$ are the tip's and
sample's density of states.
$x$ and $y$ have units of $k_V^{-1}$. }
\label{fig4}
\end{figure}
\begin{figure}
\caption{Two-protrusion STM differential conductance for an impurity
$d$-state in the same units as Fig. 2.
The two
tips are assumed to be the same distance from the impurity.
 The
differential conductance is plotted as a function of
tip-impurity distance $r$ (units of $k_V^{-1}$) and relative angle $\theta$.
 The two-protrusion STM
differential conductance shows clear four-fold angular symmetry.
The
single-protrusion STM signal corresponds to $\theta=0$.} \label{fig5}
\end{figure}
\begin{figure}
\caption{Comparison of two-protrusion
and single-protrusion STM differential conductance for a fixed tip
separation of $7.5k_V^{-1}$, where the two protrusions are oriented as
shown in Figure 1.}
\end{figure}
 \begin{figure} \caption{Differential conductance in
position space, same units as Fig. 2.
The sample is a
superconductor with a $d_{x^2-y^2}$ gap,
$eV = 0.1\Delta_{max}$,
$\Delta_{max} = 0.1\epsilon_F$, and $k_V\sim 1\AA^{-1}$. The
inelastic mean free path is taken to be $\ell_I = 500$\AA.}
\label{fig2} \end{figure} \begin{figure}
\caption{Same as Fig. 5, but for $eV=1.1\Delta_{max}$. The features are
now rotated 45$^o$ relative to Fig. 5.}
\label{fig3}
\end{figure}

\begin{references}
\bibitem{11} We assume an isotropic electron gas and
that the hopping energy from an extended state to the
localized state is proportional to their overlap.
\bibitem{12} A linear geometry has been assumed
for the tip region joining the two protrusions.
\end{references}
\end{document}